\documentclass[sigconf]{acmart}
\usepackage{multirow}
\usepackage{float}

\usepackage{todonotes} 

\def\BibTeX{{\rm B\kern-.05em{\sc i\kern-.025em b}\kern-.08emT\kern-.1667em\lower.7ex\hbox{E}\kern-.125emX}}
    
\setcopyright{none}
\renewcommand\footnotetextcopyrightpermission[1]{}
\begin{document}

%
\title{Query Expansion for Cross-Language Question Re-Ranking}

%

\author{Muhammad Mahbubur Rahman \quad Sorami Hisamoto \quad Kevin Duh}
\email{mahbubur@jhu.edu, s@89.io, kevinduh@cs.jhu.edu}
\affiliation{%
  \institution{Johns Hopkins University}
  \city{Baltimore}
  \state{MD}
  \country{USA}
}

%

%
\begin{abstract}
Community question-answering (CQA) platforms have become very popular forums for asking and answering questions daily. 
While these forums are rich repositories of community knowledge, they present challenges for finding relevant answers and similar questions, due to the open-ended nature of informal discussions. 
Further, if the platform allows questions and answers in multiple languages, we are faced with the additional challenge of matching cross-lingual information.
In this work, we focus on the cross-language question re-ranking shared task, which aims to find existing questions that may be written in different languages. 
Our contribution is an exploration of query expansion techniques for this problem. 
We investigate expansions based on Word Embeddings, DBpedia concepts linking, and Hypernym, and show that they outperform existing state-of-the-art methods. 

\end{abstract}

%
%

%

%


%

\maketitle

\section{Introduction}
Due to the huge popularity of community question-answering (CQA) platforms, such as Quora and Stack Overflow, it has captured the attention of researchers as an area with social impact.
Users who ask questions receive quick and useful answers from these community platforms.
Since these platforms are open-ended having crowd-source nature, it is a challenging task to find relevant questions and answers. 
To get the best use of these community knowledge repositories, it is very important to find the most relevant questions and retrieve their relevant answers to a new question. 
The informal writing makes this a challenge.

To deal with the need of real applications in CQA, we focus on the task of \textit{question re-ranking.}
As defined by SemEval-2016, given (i) a new question and (ii) a large collection of existing questions in the CQA platform, \textit{rerank all existing questions by similarity to the new question}, with the hope that they may provide the answers to the new question \cite{alessandromoschitti2016semeval}. 
This addresses the challenge that there exists many ways to ask the same question, and an existing question-answer pair may already satisfy the information need.

\begin{figure}
\begin{center}
\includegraphics[scale=0.34]{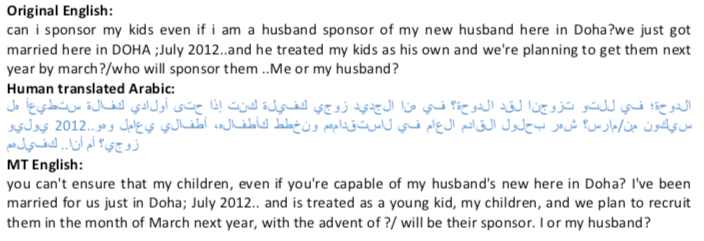}
\caption{English-Arabic-MT Question}
\label{fig:english_arabic_mt_question}
\end{center}
\end{figure}

If a CQA platform supports text entry in multiple languages, this becomes a type of cross-language information retrieval (CLIR) problem.
A machine translation (MT) model may be used to translate one language into another language prior to indexing and search, and  translation errors may lead to degradation in either precision or recall.  
For example, Fig \ref{fig:english_arabic_mt_question} shows a question in English, Arabic, and MT English from \cite{da2017cross}, who extends the SemEval-2016 task to cross-language (CL) settings: the collection of existing questions are in English, and the user queries are simulated as new questions written in Arabic.\footnote{In the original SemEval-2016 dataset, both new and existing questions were in English, but for the CL extension in \cite{da2017cross}, the new questions were replaced with their manual Arabic translations, and Arabic-to-English MT results were added to simulate a CLIR setup.}  
We can see that there are various translation errors in the MT English compared to the original English question.

\begin{figure}
\begin{center}
\includegraphics[width=0.45\textwidth]{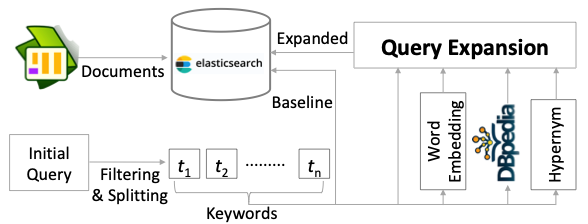}
\caption{Query Expansion Work Flow} 
[Query = Question and Documents = Existing Questions]
\label{fig:query_expansion_workflow}
\end{center}
\end{figure}

In order to address the complexity on question re-ranking in a CLIR setting for CQA platforms, we explore different query expansion (QE) techniques. 
Our hypothesis is that mis-translations are often different nuances of related concepts, so by expansion with similar terms, we may recover the underlying terms needed for matching.
We investigate Word Embedding, DBpedia, and Hypernym knowledge graph to expand query in a question-question re-ranking task. To the best of our knowledge, we first propose QE techniques for CL question re-ranking on CQA platforms. Our QE work flow is given in Fig \ref{fig:query_expansion_workflow}. 
We adopt a query translation approach, followed by QE: Given an initial query (e.g. MT English), we expand each term with information from outside resources, then match against the existing questions which are indexed as English documents in a search server like ElasticSearch. 

We develop baseline and aggregated systems using QE methods and evaluate our approaches on the CL extension of the SemEval-2016 dataset \cite{da2017cross,alessandromoschitti2016semeval}.
The evaluation results show that our QE systems achieve significant improvement over existing methods on CL question re-ranking. 



\section{Related Work}
Aiming to retrieve information in a language different from the query language, a wide range of research has been done in CLIR \cite{Nie:2010:CIR:1859417, ballesteros1996dictionary, steinberger2002cross, li2018learning, Litschko:2018:UCI:3209978.3210157, Vulic:2015:MCI:2766462.2767752}. To improve the search performance of a CLIR system, researchers have been giving more importance on QE techniques, such as, using of external lexical database \cite{miller1995wordnet}, co-occurrence analysis \cite{xu2017quary} and pseudo-relevance feedback \cite{zhai2001model, lavrenko2017relevance}. 
Zamani et al. \cite{balaneshin2017embedding}, Kuzi et al. \cite{kuzi2016query} and Diaz et al. \cite{diaz2016query} presented QE techniques using Word Embedding. A different approach using external knowledge base were developed by Xiong et al. \cite{xiong2015query} and Zhang et al. \cite{zhang2016xknowsearch} to expand queries in CLIR. 

Although, CQA \cite{bian2008finding, xiang2017answer, carmel2014improving, ji2012question, li2012analyzing} is a popular research area, there has not been much work done in CL CQA task. The CL CQA heavily depends on CL question ranking. SemEval-2016 introduced a shared task on CQA in English and Arabic \cite{alessandromoschitti2016semeval}. They also had a subtask question-question similarity ranking \cite{franco2018uh}. A later work was done by Martino et al. \cite{da2017cross} in CL question re-ranking using a CL tree kernel. There is still a big scope of improving the performance of community question re-ranking task. To the best of our knowledge, no research has been done on QE for community question re-ranking task. 




\section{Approach}

\subsection{Query Expansion}
Query Expansion is a technique of improving the search performance in IR by augmenting the initial query using different methods. To achieve a better search performance, it is important to expand the initial query to retrieve more relevant documents. Specially it is very useful CL settings since the initial query may miss important clue to retrieve more relevant documents in a different language than the original query. To expand any query we use different expansion techniques and take the union of expanded terms. For any query Q the expanded query is given in the equation \ref{eq_queryexpansion}.
\begin{equation}
  QE^{Q} = kw^{Q}  \bigcup \hspace{0.1cm} qe_{we}^{Q}  \bigcup \hspace{0.1cm} qe_{db}^{Q} \bigcup \hspace{0.1cm} qe_{h}^{Q}
  \label{eq_queryexpansion}
\end{equation}
Where $\bigcup$ is a union operator. \begin{math}QE^{Q} \end{math} is the union of all expanded queries along with the original keyword query \begin{math} kw^{Q} \end{math}. And \begin{math} qe_{we}^{Q} \end{math}, \begin{math} qe_{db}^{Q} \end{math} and \begin{math} qe_{h}^{Q} \end{math} are QE using Word Embedding, DBpedia concepts linking and Hypernym respectively.  

\subsubsection{Word Embedding}


Word embedding is a type of word representation that transforms a text into a numerical vector of real numbers where semantic similar words are grouped together. By the characteristics of word embedding, two words having semantic similar meaning share a fair amount of surrounding context. It implies that search engine should return relevant documents to the query term using the similar word. In a CL setting, due to MT, the translated text may miss the original word, and produce a different word of similar meaning, the indexed documents may not have the exact keyword. Hence the retrieval engine won't return documents of out of vocabulary.


From these intuitions and inspired by \cite{balaneshin2017embedding, kuzi2016query}, we use a pre-trained word embedding GloVe vectors \cite{pennington2014glove} based on Wikipedia and Gigaword corpus to get the most similar words for each of the query terms from the original query. The dimension of the vector is 100 and it contains 400k vocab. For a query Q having query terms \begin{math} q_1 \end{math},  \begin{math} q_2 \end{math} ......  \begin{math} q_n \end{math}, let \begin{math} t_{1}^{q_i} \end{math} and \begin{math} t_{2}^{q_i} \end{math} are the 2 most similar terms for query term \begin{math} q_i \end{math} obtained from the pre-trained word embedding model, then QE using word embedding is computed by the equation \ref{equation_word_embedding_qe}.
\begin{equation} 
    qe_{we}^{Q} = \bigcup_{i = 1}^{n}\{t_{1}^{q_i},t_{2}^{q_i}\}
    \label{equation_word_embedding_qe}
\end{equation}

As an example, a query term "travel" from a query, \textit{Im likely to travel in the month of june... just wanna know some good places to visit....}, is expanded using "travelers" and "trips" where "travelers" and "trips" are 2 most similar terms for the query term "travel".


\subsubsection{DBpedia}
DBpedia \footnote{https://wiki.dbpedia.org/} is a structured open knowledge base derived from the Wikipedia. The current version of DBpedia has 4.58 million English concepts and 38.3 millions of concepts in 125 different languages. Motivated by the work presented in \cite{xiong2015query} to associate a term with entities from Freebase knowledge graph, we use DBpedia to extract concepts and linked each query term with DBpedia concepts. 
Given a query Q having query terms \begin{math} q_1 \end{math},  \begin{math} q_2 \end{math} ......  \begin{math} q_n \end{math}, we retrieve DBpedia concepts for each of the query terms \begin{math} q_i \end{math}. Each of the returned concepts is associated with different types and properties that reflect the concept. In the expansion term selection, we choose a simple but useful strategy. We select a property called \begin{math} dct:subject \end{math} that links a concept with relevant subjects. We use the relevant subjects to expand the query term, \begin{math} q_i \end{math}. The intuition is that the concepts associated with query terms are able to capture more relevant documents. The QE is computed by the equation \ref{euation_qe_db}. 
\begin{equation}
    qe_{db}^{Q} = \bigcup_{i = 1}^{n}\bigcup_{j=1}^{k}\{ subject:q_i^j \}
    \label{euation_qe_db}
\end{equation}

Where k is the number of subjects for a query term \begin{math} q_i \end{math} and \begin{math} q_i^j \end{math} is a subject of a concept associated with the query term \begin{math} q_i \end{math}. 

As an example, a query term "travel" from a query, \textit{Im likely to travel in the month of june... just wanna know some good places to visit....}, is expanded using "Tourism", "Tourist activities" and "Transport culture" where "Tourism", "Tourist activities" and "Transport culture" are the subjects of a concept associated with "Travel". 

\subsubsection{Hypernym}
Hyponym is a specific word or phrase of a more generic term. The generic term is called hypernym. Due to MT, the translated term may have a different form of the original term. We use a publicly available hypernym knowledge graph\footnote{http://webisa.webdatacommons.org/} developed by \cite{seitner2016large} to extract hyponym labels with a high confidence score for each of the query terms and include them in the QE. The motivation is to retrieve more relevant documents which may contain hyponym terms of an original query term. For a query Q having \begin{math} q_1 \end{math}, \begin{math} q_2 \end{math} ...\begin{math} q_i \end{math}...  \begin{math} q_n \end{math} query terms, we expand Q using the equation \ref{eq_qe_hypernym}.
\begin{equation}
    qe_{h}^{Q} = \bigcup_{i =1}^{n}{\bigcup_{j=1}^{k}\{ hyponym: q_i^j \mid cs(q_i^j, q_i) >= 0.75  \}}
    \label{eq_qe_hypernym}
\end{equation}

Where, k is the number of hyponym labels for a query term \begin{math} q_i \end{math}, \begin{math} q_i^j \end{math} is a hyponym label and \begin{math} cs(q_i^j, q_i) \end{math} is a function that gives a confidence score for \begin{math} q_i^j \end{math} with respect to \begin{math} q_i \end{math}. 

As an example, a query term "travel" from a query, \textit{Im likely to travel in the month of june... just wanna know some good places to visit....}, is expanded using "operating expense", "related expense" and "personal expense" where "operating expense", "related expense" and "personal expense" are hyponym labels for "travel". 


\subsection{Search and Ranking}
To index documents, search queries and rank retrieved documents, we use Elasticsearch \footnote{https://www.elastic.co/products/elasticsearch}, a Lucene based distributed, RESTful search and analytics engine. We use a built-in English analyzer, which is used to converting documents into tokens to the inverted index for searching. The same analyzer is applied to the query string at the search time. As a ranking or scoring algorithm, we use BM25 similarity algorithm. We also configure the similarity algorithm by hyper-parameter tuning. 

\section{Dataset}
To evaluate our systems, we use SemEval-2016 Task 3 CQA dataset (Subtask B) \cite{alessandromoschitti2016semeval} and MT version of human translated Arabic questions \cite{da2017cross}. 
In \cite{da2017cross}, a new question is written in Arabic, and the goal is to retrieve similar questions in English; an MT system was available for translating the new Arabic question into English. In this research, the new question is considered as a query and a set of the first ten existing questions are considered as documents. The task is to re-rank the documents using different QE techniques.  
The dataset has 267 train, 50 dev and 70 test queries; there are 10 existing questions to be re-ranked for each new question, leading to 2670 train, 500 dev and 700 test "documents". In this research, we use only the dev and test datasets. To setup a CL environment, we choose MT version of 50 dev and 70 test questions from \cite{da2017cross} and consider them as machine translated queries. 


\section{Experimental Setup}
The datasets explained in the previous section were used for our experiments in the cross-language question re-ranking task. Initially we indexed all existing questions, which are considered as documents in this research, using the approach described in the Search and Ranking section. We observed that both English and machine translated queries might have punctuation and common words. To build a more clean queries, we filtered out punctuation and common English words from the initial queries. Then we split each query into words which are considered as keyword query, our baseline system. We experimented in two scenarios: a) English query, the original SemEval-2016 task B, and b) Machine translated query, where Arabic version of the English queries are translated back into English. 
The second scenario is the CLIR setting we focus here, while the first scenario provides comparison results in the monolingual setting. 

We configured 18 systems for English and MT queries, by combining each of the four basic systems: (a) Keyword, (b) Word Embedding, (c) DBpedia and (d) Hypernym. The system (a) is the baseline system whereas (b), (c) and (d) are systems based on QE using Word Embedding, DBpedia and Hypernym knowledge graph respectively. The average query lengths for baseline systems are 18.67(EN) and 19.96(MT) words. And the average word additions are Word Embedding: 30.98(EN) and 35.34(MT); DBpedia: 25.81(EN) and 31.32(MT); Hypernym: 21.58(EN) and 29.04(MT); Best system: 78.3(EN) and 95.6(MT). 
The combination of 18 systems are given in table \ref{table:mapscores_all}. All the systems are experimented in two settings - \textbf{Dev} and \textbf{Test}. The search ranking scores are calculated for all 18 systems using BM25 algorithm. We tuned BM25 hyper-parameters, k1 and b on \textbf{Dev} set to get the optimized values where k1 controls non-linear term frequency normalization and b controls to what degree document length normalizes \begin{math} tf \end{math} values. The score for each query is calculated based on 10 existing documents to re-rank them. 


\begin{table}[t]
\begin{tabular}{llccc}
\hline
\multicolumn{1}{|c|}{\textbf{System}} & \multicolumn{1}{c|}{\textbf{QR}} & \multicolumn{2}{c|}{\textbf{MAP}} & \multicolumn{1}{|c|}{\textbf{\begin{math}\Delta\end{math}}}                                      \\ \cline{3-4} 
\multicolumn{1}{|c|}{}  & \multicolumn{1}{c|}{} & \multicolumn{1}{c|}{\textbf{Dev}} & \multicolumn{1}{c|}{\textbf{Test}} & 
\multicolumn{1}{|c|}{\textbf{}} \\ \hline\hline
\multicolumn{1}{|l|}{1. Keyword(KW) (Baseline)}                                  & \multicolumn{1}{l|}{EN}                                              & \multicolumn{1}{c}{72.60}                                 & \multicolumn{1}{c|}{71.43} & \multicolumn{1}{l|}{\hspace{0.2cm}00.00}                                \\
\multicolumn{1}{|l|}{2. Word Embedding(WE)}                                      & \multicolumn{1}{l|}{EN}                                              & \multicolumn{1}{c}{64.40}                                 & \multicolumn{1}{c|}{63.86} & \multicolumn{1}{l|}{\hspace{0.1cm}-07.57}                                 \\
\multicolumn{1}{|l|}{3. DBpedia(DB)}                                            & \multicolumn{1}{l|}{EN}                                              & \multicolumn{1}{c}{41.00}                                 & \multicolumn{1}{c|}{45.29} & \multicolumn{1}{l|}{\hspace{0.1cm}-26.14}
                      \\
\multicolumn{1}{|l|}{4. Hypernym(HN)}                                        & \multicolumn{1}{l|}{EN}                                                 & \multicolumn{1}{c}{21.00}                                 & \multicolumn{1}{c|}{27.86} & \multicolumn{1}{l|}{\hspace{0.1cm}-34.57}
                \\
\multicolumn{1}{|l|}{5. 1 + 2 (KW+WE)}                                              & \multicolumn{1}{l|}{EN}                                             & \multicolumn{1}{c}{80.20}                                 & \multicolumn{1}{c|}{79.86} & \multicolumn{1}{l|}{\hspace{0.1cm}+08.43}                                 \\
\multicolumn{1}{|l|}{6. 1 + 3 (KW+DB)}                                             & \multicolumn{1}{l|}{EN}                                              & \multicolumn{1}{c}{76.00}                                 & \multicolumn{1}{c|}{75.29} & \multicolumn{1}{l|}{\hspace{0.1cm}+03.86}                                 \\
\multicolumn{1}{|l|}{7. 1 + 4 (KW+HN)}                                              & \multicolumn{1}{l|}{EN}                                             & \multicolumn{1}{c}{75.20}                                 & \multicolumn{1}{c|}{76.00}  & \multicolumn{1}{l|}{\hspace{0.1cm}+04.57}                                \\
\multicolumn{1}{|l|}{8. 2 + 3 + 4 (WE+DB+HN)}                                              & \multicolumn{1}{l|}{EN}                                             & \multicolumn{1}{c}{72.20}                                 & \multicolumn{1}{c|}{75.86}   & \multicolumn{1}{l|}{\hspace{0.1cm}+04.43}                               \\
\multicolumn{1}{|l|}{\textbf{9. 1 + 2 + 3 + 4 (Best)}}              & \multicolumn{1}{l|}{EN}                                              & \multicolumn{1}{c}{\textbf{84.00}}                                 & \multicolumn{1}{c|}{\textbf{82.00}} &\multicolumn{1}{l|}{\hspace{0.1cm}+10.57}
              \\
\multicolumn{1}{|l|}{10. UH-PRHLT(SemEval\cite{alessandromoschitti2016semeval,franco2018uh})}                               & \multicolumn{1}{l|}{EN}                                              & \multicolumn{1}{c}{75.90}                                 & \multicolumn{1}{c|}{76.70}   & \multicolumn{1}{l|}{\hspace{0.4cm}-}          \\
\multicolumn{1}{|l|}{11. SVM + TK \cite{da2017cross}}                              & \multicolumn{1}{l|}{EN}                                              & \multicolumn{1}{c}{73.02}                                 & \multicolumn{1}{c|}{77.41}  & \multicolumn{1}{l|}{\hspace{0.4cm}-}                                \\\hline\hline
\multicolumn{1}{|l|}{12. Keyword(KW) (Baseline)}                                 & \multicolumn{1}{l|}{MT}                                                  & \multicolumn{1}{c}{72.20}                                 & \multicolumn{1}{c|}{67.57}  & \multicolumn{1}{l|}{\hspace{0.2cm}00.00}                                \\
\multicolumn{1}{|l|}{13. Word Embedding(WE)}                                    & \multicolumn{1}{l|}{MT}                                                  & \multicolumn{1}{c}{64.40}                                 & \multicolumn{1}{c|}{63.43}  & \multicolumn{1}{l|}{\hspace{0.1cm}-04.14}                                \\
\multicolumn{1}{|l|}{14. DBpedia(DB)}                                        & \multicolumn{1}{l|}{MT}                                                      & \multicolumn{1}{c}{43.20}                                 & \multicolumn{1}{c|}{45.71}     & \multicolumn{1}{l|}{\hspace{0.1cm}-21.86}                             \\
\multicolumn{1}{|l|}{15. Hypernym(HN)}                                          & \multicolumn{1}{l|}{MT}                                                  & \multicolumn{1}{c}{26.80}                                 & \multicolumn{1}{c|}{32.71} & \multicolumn{1}{l|}{\hspace{0.1cm}-34.86}                                 \\
\multicolumn{1}{|l|}{16. 12 + 13 (KW+WE)}                                            & \multicolumn{1}{l|}{MT}                                                  & \multicolumn{1}{c}{79.20}                                 & \multicolumn{1}{c|}{75.57}    & \multicolumn{1}{l|}{\hspace{0.1cm}+08.00}                              \\
\multicolumn{1}{|l|}{17. 12 + 14 (KW+DB)}                                            & \multicolumn{1}{l|}{MT}                                                  & \multicolumn{1}{c}{75.40}                                 & \multicolumn{1}{c|}{71.43}   & \multicolumn{1}{l|}{\hspace{0.1cm}+03.86}                               \\
\multicolumn{1}{|l|}{18. 12 + 15 (KW+HN)}                                            & \multicolumn{1}{l|}{MT}                                                  & \multicolumn{1}{c}{76.40}                                 & \multicolumn{1}{c|}{72.29}      & \multicolumn{1}{l|}{\hspace{0.1cm}+04.72}                            \\
\multicolumn{1}{|l|}{19. 13 + 14 + 15 (WE+DB+HM)}                                            & \multicolumn{1}{l|}{MT}                                                  & \multicolumn{1}{c}{77.60}                                 & \multicolumn{1}{c|}{73.14}   & \multicolumn{1}{l|}{\hspace{0.1cm}+05.57}                               \\

\multicolumn{1}{|l|}{\textbf{20. 12 + 13 + 14 + 15(Best)}}           & \multicolumn{1}{l|}{MT}                                                  & \multicolumn{1}{c}{\textbf{84.00}}                                 & \multicolumn{1}{c|}{\textbf{78.29}} & \multicolumn{1}{l|}{\hspace{0.1cm}+10.72}                                 \\
\multicolumn{1}{|l|}{21. SVM+TK(\cite{da2017cross})}                              & \multicolumn{1}{l|}{MT}                                                  & \multicolumn{1}{c}{72.94}                                 & \multicolumn{1}{c|}{76.67}  & \multicolumn{1}{l|}{\hspace{0.4cm}-}                                \\\hline

\end{tabular}

\caption{
MAP scores for various QE on English (EN) questions (monolingual setup) and MT questions (CLIR setup). 
}
\label{table:mapscores_all}
\end{table}

\begin{figure*}[t]
\begin{center}
\includegraphics[height=1.2in, width=1.00\textwidth]{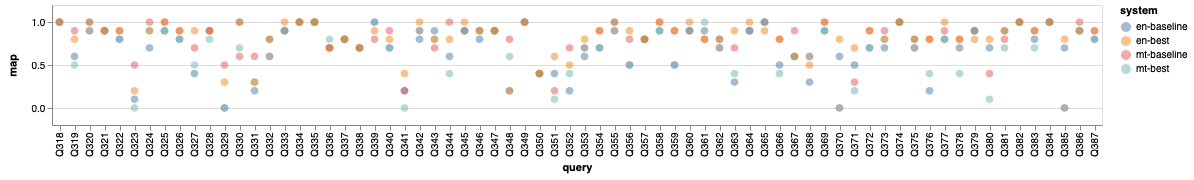}
\caption{MAP scores over queries}
\label{fig:map_scatter_plot}
\end{center}
\end{figure*}

\section{Results and Analysis}
Table \ref{table:mapscores_all} compares the MAP scores on the dev and test sets. The "System" column compares QE, Keyword baseline without QE, and previously-published state-of-the-art methods in this task (UH-PRHLT, SVM+TK); the QR column shows whether the query was English (monolingual setup) or Machine-translated English (cross-language setup).
The \begin{math} \Delta \end{math} column displays the difference between test MAP against the baseline.

The results on the original English queries are shown in rows 1 to 11. Row 1 is the baseline system using keyword query without any QE. Rows 2 to 4 show scores for QE using Word Embedding, DBpedia and Hypernym respectively. We observe that each of the individual systems from rows 2 to 4 has a negative \begin{math} \Delta \end{math} score. That means QE using any single approach doesn't beat the baseline system. Combination of baseline and any QE method are shown in rows 5 to 7 where each of the combinations has better performance than the baseline system. Among them, combination of word embedding and the baseline system has the highest \begin{math} \Delta \end{math} score which is +08.43. Row 8 shows QE using a combination of Word Embedding, DBpedia and Hypernym, which also beats the baseline system by +04.43 \begin{math} \Delta \end{math} score. An aggregation of all systems from rows 1 to 4 is shown in row 9, which is the best performing system. The best system beats our baseline system by +10.57 \begin{math} \Delta \end{math} score which is a substantial improvement over the baseline system. 

The best performing systems from the SemEval-2016 Task 3 \cite{alessandromoschitti2016semeval, franco2018uh} and the state-of-the-art system \cite{da2017cross} in the community question re-ranking task are shown in row 10 and 11 respectively. Our best system outperforms them by +05.30 and +04.59 MAP scores respectively which is an effective improvement in the community question re-ranking task.

The results on MT queries are presented in rows 12 to 21 where the baseline system is shown in row 12. Individual QE using Word Embedding, DBpedia and Hypernym are displayed in rows 13 to 15. We find a similar pattern between these and the QE in English that is negative \begin{math} \Delta \end{math} score which implies individual QE technique also doesn't perform well for MT queries. Union of the baseline system and any other expansion method from rows 13 to 15 are given in rows 16 to 18. Similar to the English query, we also achieve positive \begin{math} \Delta \end{math} scores for each of them where expansion using Word Embedding has the highest \begin{math} \Delta \end{math} score +08.00.

Row 19 shows a combination of QE methods which also beats the baseline by +05.57 MAP score. The best system, union of baseline and all QE given in row 20, improves the performance by +10.72 \begin{math} \Delta \end{math} score. Our best system in MT setting also outperforms the state-of-the-art system given in row 21 by +01.62 MAP score. The significant improvement compared to the baseline and the state-of-the-art, implies that our QE approaches are also strong to MT.

In the comparison of baseline systems in English and MT (row 1 and 12), we notice that the MT baseline system has a lower MAP score by -3.86. We also observe that the MT best system degrades the MAP score by -03.71 than that of English. We assume the reason behind these low map scores for MT systems is the effect of the output of machine translation. We see that individual QE using DBpedia and Hypernym have slightly better performance in MT than English by +00.42 (diff. between row 14 and row 3) and +04.85 (diff. between row 15 and row 4) map scores respectively. 

Most importantly, we find that our best systems (English and MT), outperform the baseline systems and state-of-the-art results in community question re-ranking task. These indicate that our QE methods are robust and effective in both monolingual and CL settings. Figure \ref{fig:map_scatter_plot} shows MAP scores for each query of baseline and best systems for both English and MT. One interesting observation is that we get MAP score 0 for 5 out of 70 test queries, and all these are for either MT baseline or MT best system. Only query 329 has a 0 MAP for English baseline system along with MT best system. 
\section{Conclusion}
We investigate different query expansion techniques for improving cross-language question re-ranking in community question answering platforms. Our techniques, though simple, outperform current state-of-the-art on SemEval-2016 Task 3 and its CLIR extension.
As a future work, we plan to improve methods for candidate terms selection for each of the different query expansions types.

%


\begin{acks}
This work is supported in part by the Office of the Director of National
Intelligence (ODNI), Intelligence Advanced Research Projects Activity
(IARPA), via contract \#FA8650-17-C-9115. The views and conclusions
contained herein are those of the authors and should not be interpreted
as necessarily representing the official policies, either expressed or
implied, of ODNI, IARPA, or the U.S. Government. The U.S. Government is
authorized to reproduce and distribute reprints for governmental
purposes notwithstanding any copyright annotation therein. 
\end{acks}

%
\bibliographystyle{ACM-Reference-Format}
\bibliography{references}


\end{document}